%% file: llm-dse.tex
\documentclass{article}

\usepackage[preprint]{neurips_2025}

\usepackage[utf8]{inputenc} 
\usepackage[T1]{fontenc}    
\usepackage{hyperref}       
\usepackage{url}            
\usepackage{booktabs}       
\usepackage{amsfonts}       
\usepackage{nicefrac}       
\usepackage{microtype}      
\usepackage{makecell}
\input{macro}

\title{LLM-DSE: Searching Accelerator Parameters with LLM Agents}

\author{%
  Hanyu Wang\textsuperscript{*}\textsuperscript{\dag}
  \\
  \And
  Xinrui Wu\textsuperscript{*}\textsuperscript{\dag}\\
  \And
  Zijian Ding\textsuperscript{*}\textsuperscript{\dag}\\
  \And
  Su Zheng\textsuperscript{‡}\\
  \And
  Chengyue Wang\textsuperscript{\dag}\\
  \And
  Neha Prakriya\textsuperscript{\dag}\\
  \And
  Tony Nowatzki\textsuperscript{\dag}\\
  \And
  Yizhou Sun\textsuperscript{\dag}\\
  \And
  Jason Cong\textsuperscript{\dag}\\
}

\begin{document}

\maketitle

\begingroup
\renewcommand\thefootnote{}\footnotetext{\textsuperscript{*}Equal contribution}
\renewcommand\thefootnote{}\footnotetext{\textsuperscript{\dag}UCLA}
\renewcommand\thefootnote{}\footnotetext{\textsuperscript{‡}CUHK}
\endgroup

\begin{abstract}
Even though high-level synthesis (HLS) tools mitigate the challenges of programming domain-specific accelerators (DSAs) by raising the abstraction level, optimizing hardware directive parameters remains a significant hurdle. Existing heuristic and learning-based methods struggle with adaptability and sample efficiency.
We present LLM-DSE, a multi-agent framework designed specifically for optimizing HLS directives. Combining LLM with design space exploration (DSE), our explorer coordinates four agents: Router, Specialists, Arbitrator, and Critic. These multi-agent components interact with various tools to accelerate the optimization process.
LLM-DSE leverages essential domain knowledge to identify efficient parameter combinations while maintaining adaptability through verbal learning from online interactions. Evaluations on the HLSyn dataset demonstrate that LLM-DSE achieves substantial $2.55\times$ performance gains over state-of-the-art methods, uncovering novel designs while reducing runtime. Ablation studies validate the effectiveness and necessity of the proposed agent interactions. Our code is open-sourced here: \href{https://github.com/Nozidoali/LLM-DSE}{LLM-DSE}.
\end{abstract}

\section{Introduction}

In recent decades, due to the increasing computational demands in both data centers and edge devices, domain-specific accelerators (DSAs) have emerged across various application domains~\citep{2018ieee,dally2020domain}. For example, DSAs have been developed for critical algorithms in genomic sequencing like Minimap2~\citep{li2018minimap2,guo2019hardware,turakhia2018darwin}, and are also widely adopted to enable scientific research in high-energy physics~\citep{wirthlin2015high} and quantum computing~\citep{acharya2024quantum}. They have also played a pivotal role in speeding up the training and inference of modern deep neural networks~\citep{tpu2017,he2025intrra}. 

However, designing accelerators remains a significant challenge~\citep{overgen}, limiting their accessibility to hardware experts. Our goal in this work is to study whether modern AI techniques, coupled with advanced toolchains, could make developing DSAs more accessible to all researchers, which will in turn boost their process of making scientific discoveries.

\begin{figure*}[t]
    \centering
    \includegraphics[width=0.95\textwidth]{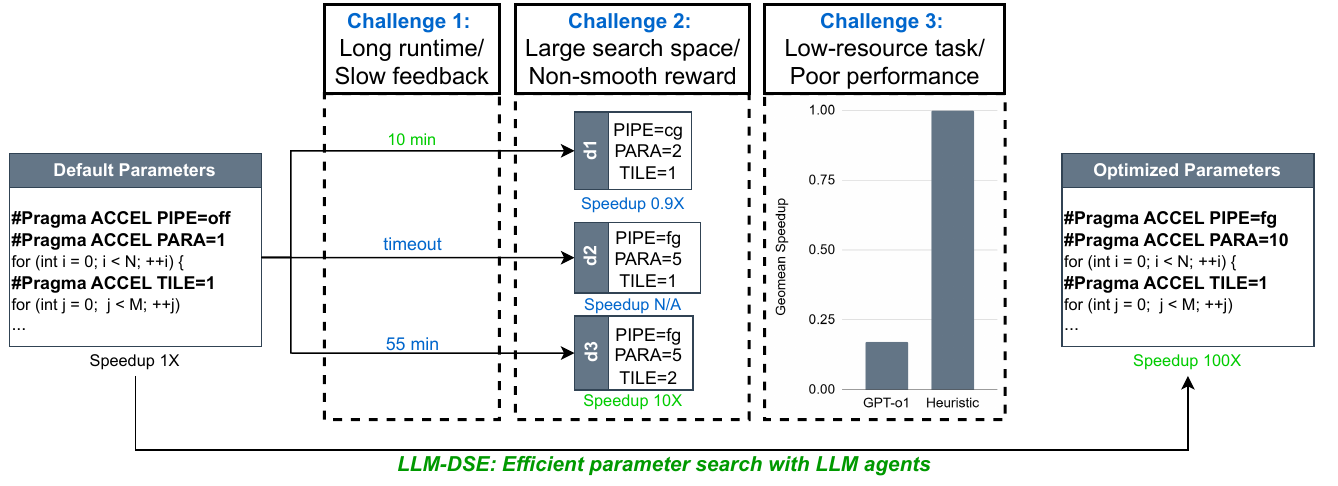}
    \caption{Searching accelerator parameters is a challenging task.}
    \label{fig:motivation}
\vspace{-1em}
\end{figure*}

High-level synthesis (HLS) alleviates the difficulty of programming DSAs by raising the abstraction level from register-transfer level (RTL) design to C/C++ (Appx. \ref{appx:hls}). As illustrated in Figure~\ref{fig:motivation}, generating efficient hardware depends on selecting the appropriate combination of parameters.

Despite efforts to simplify hardware design, HLS tools and design frameworks such as Merlin Compiler~\citep{merlin} need an efficient parameter search strategy. Indeed, the parameter space for hardware optimization is often vast ($\approx10^{13}$), and an evaluation of each design point requires running the downstream synthesis flow (takes hours). Therefore, an exhaustive search is impractical. 

Existing solutions are categorized into heuristic-based~\citep{autodse} and model-based~\citep{harp} for automatic parameter tuning. However, both approaches suffer from poor generalization when applied to new kernels or configurations. Heuristic-based methods struggle to find optimal configurations as no single heuristic is universally effective across all workloads. Model-based methods rely on large data sets to train surrogate models of the tool behavior, yet collecting these datasets is time-consuming due to the long runtime required to evaluate each design point. 

This work investigates the use of large-language model (LLM)-based agents to optimize DSA parameters, particularly in a data-efficient manner (Figure \ref{fig:motivation}). Compared to state-of-the-art search methods, our framework follows domain-specific instructions to refine the optimization parameters while maintaining feasibility, interacts with various tools to incorporate feedback and enhance design processes, and dynamically adjusts strategies based on real-time insights (Figure \ref{fig:llm-dse-workflow}).

More specifically, the core of LLM-DSE is to inject domain knowledge into an iterative tree-search process. The search process involves selecting parent configurations from the searching frontier, expanding child configurations from the selected parents, selective evaluation of newly expanded children, and branch pruning. These four steps are implemented as different LLM agents, resulting in an iterative refinement process powered by specialized components that coordinate global and local decisions. Given the heterogeneity of different types of hardware parameters, a multi-agent approach is necessary, with each agent focusing on a specific parameter type to enhance design efficiency.

Equipped with domain-specific instructions and adaptive strategies, LLM-DSE designed accelerators demonstrate an average $2.5\times$ speedup over heuristic-based methods and $6\times$ over model-based approaches on workloads derived from the HLSyn dataset~\citep{hlsyn} from the second stage of the ML4HLS contest~\citep{hlscontest}. Evaluations on the additional four Rosetta benchmarks~\citep{rosetta} further demonstrate the scalability of our approach to larger programs, achieving a geometric mean speedup of $1.22\times$ compared to the heuristic-based method. We also show that our approach could generalize to different toolchains with minimal changes in the prompts~(Appx. \ref{appx:hls}, \ref{appx:multi_bknd}). 

\section{Searching Accelerator Parameters is a Challenging Task}
HLS directive optimization can be formulated as a constrained optimization problem. 
A design point $\mathbf{d}$ is encapsulated by a set of directives, $\mathbf{d} = (d_1, d_2, \dots, d_n)$, that applies to a software program $p$. 
Different parameter configurations lead to hardware implementations with varying performance and resource consumption. Our objectives are:

\textbf{Performance optimization}. We minimize the latency in clock cycles to complete the program on our hardware accelerator. 

\textbf{Honor resource constraints}. We keep utilization of hardware resources (e.g., LUTs, BRAM, FFs, DSPs, and URAM) to be less than $80\%$, and save enough resources for additional Input/Output modules.

Note that the downstream automated synthesis, verification, and refinement tools are correct by construction during source-to-source translation~\citep{merlin}. Therefore, we honor the functionality constraints by constructing a set of design rules. More specifically, we require that the configuration $\bold{d}$ lies within the unique optimization space $\mathcal{D}_p$ associated with each program $p$. 

With downstream tools, design rules suffice to ensure the feasibility and correctness of our hardware accelerator. However, three challenges remain:

\textbf{Challenge 1: Slow feedback.} HLS flow reveals latency and resource usage only after all the downstream flows have finished. Obtaining ground-truth feedback for every parameter setting is, therefore, time-consuming, and for aggressively optimized configurations, the flow may even fail to finish, leaving no feedback at all.

\textbf{Challenge 2: Huge, non-smooth search space.} The search space is huge and the reward is not smooth. For example, as illustrated in Figure~\ref{fig:motivation}, modifying the parameters from d2 to d3 can change the outcome from ``timeout'' to a highly performant design. Moreover, efficient parameter combinations may take more time to synthesize, resulting in a more complex reward landscape near the high-performing regions. 

\textbf{Challenge 3: Low-resource setting.} The task is low-resource, resulting in poor direct generation performance in Figure~\ref{fig:motivation} (details in \textsection{} \ref{exp:generation}). 


\section{Method}\label{sec:method}

We propose \textbf{LLM-DSE}, where hardware accelerator parameter searching is framed as a closed-loop, multi-agent design space exploration~(DSE) process. The tree-search process is displayed in Figure~\ref{fig:llm-dse-workflow}, where nodes stand for designs and edges represent parameter updates. 

Our search starts with the most conservative parameter assignment (without any parallelism) and improves performance progressively.  
This default design implements the kernel straightforwardly and is usually fast to compile. 
Therefore, it ensures that our exploration intersects with the feasible region and returns at least one design that honors the resource constraints. All explored designs are stored during DSE. In each iteration, LLM-DSE analyzes the exploration history and decides the next optimization (parameter update) to exploit. In the rest of this section, we illustrate three key mechanisms in LLM-DSE. 

\subsection{Routing Tasks to Appropriate Specialists}\label{subsec:specialist}
\begin{figure*}[t]
\centering
\includegraphics[width=\textwidth]{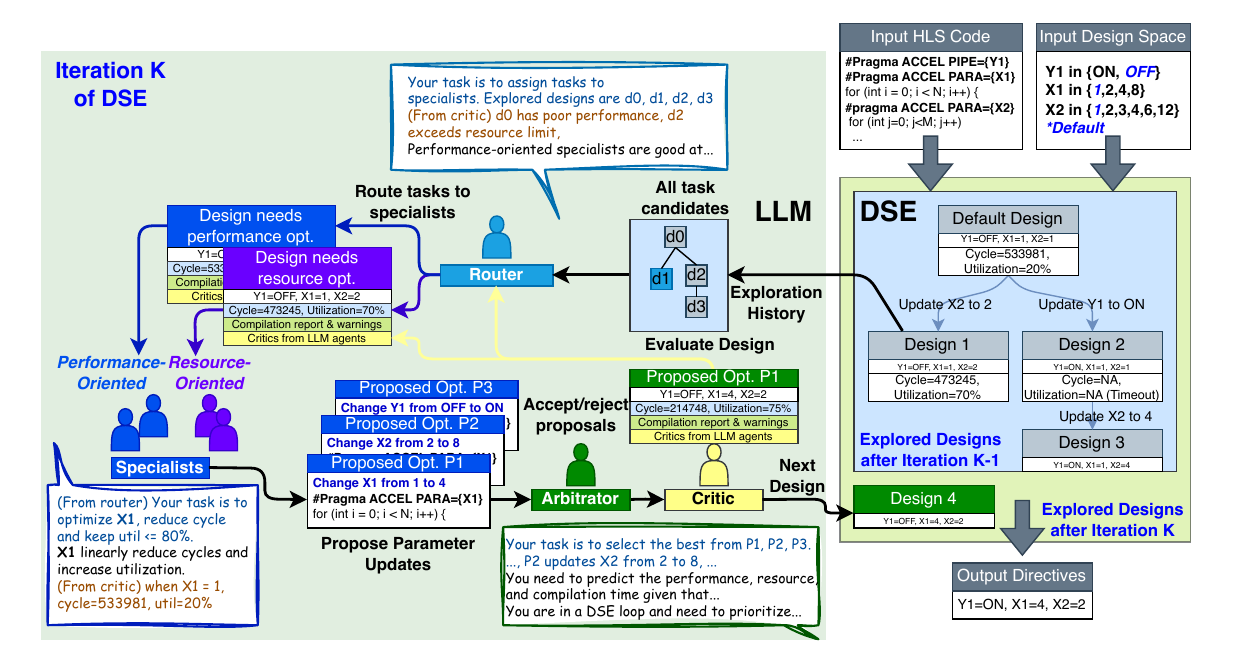}
\caption{LLM-DSE overview. The inputs are the HLS code with a set of parameters to be determined, and the design space that specifies the value ranges. The output is a design with all parameter values. It coordinates different agents and navigates in the design space via a collaborative process between the router, a group of specialists, an arbitrator, and the critic.}
\label{fig:llm-dse-workflow}
\vspace{-1em}
\end{figure*}

In each DSE iteration, our \emph{task} is to optimize a design $\mathbf{d}$ from exploration history to $\mathbf{d'}$. 
Design $\mathbf{d'}$ is more optimal than $\mathbf{d}$ if it has better performance or fixes the resource violation. 
%
%
Due to the low-resource nature of our task (Challenge 3), LLM agents cannot directly handle the whole problem and are likely to be misguided by LLMs' hallucination.
To target multiple objectives, LLM-DSE employs \emph{Specialists} with heterogeneous expertise. 
As shown in Figure~\ref{fig:llm-dse-workflow}, we employ two flavors of specialists, performance-oriented and resource-oriented, each focusing on distinct hardware accelerator design objectives.
To specialize the same LLM model (with the same weights) to different roles, we leverage the in-context learning with tailored prompts so that each specialist brings a unique perspective to parameter tuning. 

\textbf{Therefore, specialists are equipped with complementary backgrounds}. 
Performance-oriented specialists leverage given performance modeling knowledge and the history of past configurations to predict cycle count improvements for directives such as loop unrolling or pipelining. Resource regulation specialists, armed with utilization analytics and past exploration data, forecast how transformations such as tiling will impact on-chip resource usage. 
By equipping each specialist with directive-specific knowledge and historical context, we ensure that proposals are both accurate and diverse.

\textbf{Meanwhile, different parameters exhibit distinct characteristics}. 
For instance, the pipelining directive increases the performance aggressively and is best suited for configurations that have a large space for improvement. 
In contrast, the tiling directive is more effective on designs with higher resource utilization or those that experience compilation timeouts. 

To harness this diversity, LLM-DSE employs a \emph{Router} to assign design optimization tasks to these specialists. 
For each candidate design $\mathbf{d}$, it analyzes its characteristics and optimization history to determine which specialist is best suited for the task. 
While conventional heuristic-based approaches simply pick the current best configuration, LLM-DSE's router evaluates the optimization potential of each design based on feedback from the critic and assigns it to specialists based on a more robust bottleneck analysis. 

\subsection{Dual-Stage Proposal Filtering Pipeline}\label{subsec:arbitrator}
Deciding which update ($\mathbf{d}\rightarrow\mathbf{d'}$) to exploit is critical in a DSE process, particularly for hardware design, as the evaluation cost is extremely high and only limited attempts are allowed before returning the final design (Challenge~1). To make a more thorough parameter update decision, we design a dual-stage proposal filtering pipeline between the specialists and the arbitrator. 

\textbf{In stage 1}, specialists propose one value update (i.e., $d_i\rightarrow d_i'$) for each parameter $i$. Note that this value ($d_i'$) are not constrained to the immediate neighborhood of the current design ($d_i$). 
Instead, specialists can propose larger jumps such as updating the \texttt{PARA} factor \texttt{X1} from 1 to 4 (proposal P1 in Figure~\ref{fig:llm-dse-workflow}). 
This is made possible because our specialists have the capability to undo or recover the over-aggressive updates in later iterations. 


\textbf{In stage 2}, after specialists generate candidate updates, they submit them to the \emph{Arbitrator}. Instead of accepting all proposals, the arbitrator consolidates them and selects the updates more likely to accelerate convergence (e.g., proposal P1 in Figure~\ref{fig:llm-dse-workflow}). 

First, similar to the specialists, our arbitrator is equipped with knowledge about how directives would affect performance and resource utilization and is prompted to predict the results after applying each proposed update. 
However, while specialists are responsible for deciding the values of each parameter, the arbitrator focuses on the overall design and is specialized in comparing different parameters' sensitivity.

Second, the arbitrator is aware of the remaining exploration budget. Depending on the stage, the agent tends to aggressively optimize the objectives with higher uncertainties at early stages and becomes relatively conservative when selecting candidates in later iterations. 

This dual-stage filtering process further enhances the exploration efficiency. 
Although the specialists are instructed to provide accurate and diverse suggestions, they may not always agree on the best course of action. 
Due to the expensive nature of evaluating each design, it is crucial to select only a few proposals for evaluation. 
LLM-DSE's arbitrator is designed to make this final decision by balancing the trade-off between exploration and exploitation.

\subsection{Handling Context Explosion and Vague Reports with Tools}
\textbf{We extract feedback from report and log}. After accepted by the arbitrator, the new design $\mathbf{d'}$ will be evaluated by the \emph{Critic}. To achieve this, critic first utilizes the downstream tools to generate the compilation report and collects the warnings in the log. 
The numerical results such as cycle count and resource utilization are parsed from the report. 

Although these results may provide feedback to the router and specialists, such feedback does not necessarily correlate directly to the policies in later iterations (Challenge~2). In this case, those agents are likely to be overfitted or misled by past experiences, causing them to get stuck at a local optimum branch without making progress.

To mitigate this issue, critic also performs a comparison-based branch pruning and evaluates the parameter updates and eliminates unpromising branches in the search space. Note that the comparison is between the child $\mathbf{d}'$ and its corresponding parent $\mathbf{d}$, between which only one directive $d_i$ is different. 
This decouples the influence of different parameters in the comparison and improves the accuracy of the judgment. Based on the comparison result, the critic generates natural-language feedback on the explored configurations and candidate branches. This feedback will be added in the prompts for the router and specialists agents, enabling a more direct channel to pass branch-pruning messages.



\textbf{However, we must handle context explosion.} As LLM-DSE continues to explore new design parameters, the number of candidate tasks for the router grows linearly. Appending all available tasks to the router's context can cause context explosion or harm the fidelity of its decision. We alleviate this issue by introducing a tool, \emph{history curator}, that selects only top $K$ most representative designs based on both performance and parameter assignment diversity (i.e., filtering similar designs). The router is advised to call this tool after exploring enough parameters.

Another source of long context is the report given by the toolchain. As all agents rely on feedback from multiple reports to make decisions, presenting the full report to them is infeasible. We thus parse and extract only the necessary information for the agents. Our parser utilizes the structured report provided by the toolchain and handles common errors like compilation timeout. We treat the design as invalid if the compiler raises an unknown error.

Our task selection tool and report parser manage an acceptable context length throughout the parameter search. They also free agents from handling vague or incomplete reports.

\section{Experiments}
\subsection{Experimental Setup}\label{exp:setup}

\textbf{Synthesis flow details}. All accelerator designs are evaluated under the same high-level synthesis workflow Merlin~\citep{merlin_compiler} with the same tool version and hyperparameters. We evaluate the best configuration that is successfully compiled and synthesized within the specified duration. We set the search timeout to be 8 hours for Merlin compiler. For the ablations on different toolchains, we set the search timeout to 8 hours on Stratus~\citep{stratus} and to 1 hour on Vitis~\citep{vitis24} due to the significantly faster synthesis time of on-chip modules compared to Stratus's ASIC flow and Merlin's end-to-end flow.

\textbf{Evaluation benchmarks}. The benchmark suite used in our experiments consists of 10 representative workloads from HLSyn~\citep{hlsyn}. These kernels are used in the ML4HLS contest~\citep{hlscontest} with different loop structures and varying loop bound values, providing a comprehensive evaluation of our approach. 

\textbf{Implementaton details}. Our framework is implemented in Python and employs the LLM via the OpenAI API. All experiments are conducted on a system equipped with 60 cores of an AMD EPYC 7V13 CPU and 240GB of RAM.

\subsection{Comparison with Existing Heuristic- and Model-based Methods}\label{exp:baselines}

\input{tables/comparison}

\textbf{Heuristic-Based:} We compare our method with AutoDSE~\citep{autodse}, the state-of-the-art heuristic-based approach for HLS directive optimization. AutoDSE leverages extensive domain knowledge to efficiently explore the design space. Following the preference in the chip design community for prioritizing performance improvements over runtime constraints, we evaluate AutoDSE under two different runtime budgets: 8 hours and 24 hours, both using a batch size of 8 for parallel exploration. We denote these baselines as AutoDSE-8 (A8) and AutoDSE-24 (A24), respectively.  

\textbf{Model-Based:} We also compare against HARP~\citep{harp}, a model-based approach for HLS directive optimization. HARP employs a Graph Neural Network (GNN) as the surrogate model for HLS tools and performs a breadth-first search (BFS) using the surrogate model for one hour to identify promising design for evaluation. We evaluate HARP trained on datasets generated from AutoDSE in 8 hours (H8) and 24 hours (H24).

Table~\ref{tab:performance-evaluation-reordered} presents a comparison of the design discovered by our framework and the baseline methods in terms of the number of clock cycles ($\downarrow$). Each experiment is run twice, and the standard deviation is recorded. The speedup is calculated as the ratio of the baseline's cycle count to ours. On average, our design achieve speedups of $2.55\times$, $1.60\times$, and $1.16\times$ compared to AutoDSE-8, AutoDSE-24, and HARP-24, respectively.

Overall, LLM-DSE significantly outperforms AutoDSE and HARP, producing parameter configurations that complete execution in fewer cycles. As shown in Table~\ref{tab:performance-evaluation-reordered}, our framework consistently discovers superior designs across various benchmarks. In certain cases, such as \texttt{syr2k}, our method identifies high-performance designs that dominate all previously explored solutions on the same hardware platform.  

This advantage likely stems from the inherent biases introduced by heuristic-based methods, which may limit their ability to explore certain regions of the search space. Notably, neither AutoDSE nor HARP are able to reach the designs discovered by our framework, even with extended search durations. Our LLM-DSE framework facilitates more substantial leaps in the design space by allowing the specialists to update parameter values more aggressively. We conclude that our agentic workflow not only enhances performance but also reduces runtime, making it a more effective approach for HLS parameter optimization.

\subsection{Ablations}\label{exp:abl}
We perform two types of ablations. First, we study whether interactions between different agents are necessary. We compare LLM-DSE with two simpler agent workflow architectures. Next, we study whether (under the same architecture) each agent has to be LLM-based. We gradually replace each agent with fixed heuristic to validate the necessity of using LLM-based agents.

\subsubsection{Ablating Agent Interactions}
Note that the two core agent interactions in LLM-DSE are: (1) between the router and the specialists, and (2) between the specialists and the arbitrator. We develop two corresponding ablations: (1) removing the arbitrator and sending all the proposed parameter updates to the critic, and (2) keeping only the performance-oriented specialists.

\begin{wraptable}{r}{0.5\textwidth}  
  \vspace{-1.5em}                    
  \centering
  \small
  \caption{Ablation: agent interactions}\label{tab:abl_agent_arch}
  \setlength{\tabcolsep}{4pt}
  \begin{tabular}{lcc}
    \toprule
    \textbf{\makecell[l]{Speedup over\\ Simpler Archs}} & \textbf{\makecell[l]{w/o\\ arbitrator}} & \textbf{\makecell[l]{only performance-\\oriented specialists}} \\
    \midrule
    Geomean & $1.58\times$ & $1.10\times$ \\
    Win \% & $70\%$ & $50\%$ \\
    Tie \% & $10\%$ & $10\%$ \\
    \bottomrule
  \end{tabular}
  \vspace{-1em}
\end{wraptable}

\textbf{Without arbitrator.} Table \ref{tab:abl_agent_arch} presents the result when we remove the arbitrator and send all the parameter updates to the critic. We show that due to the long run-time of each evaluation, with the arbitrator, LLM-DSE can better utilize the limited budget of getting ground-truth feedback. While the critic also has the flexibility to reject new parameter updates, the arbitrator will compare the proposals of all specialists, making it's role important and unique for this combinatorial optimization task.

\textbf{Only performance-oriented specialists.} When removing resource-oriented specialists group, we also see a slight decrease in overall performance. As detailed in the Appendix \ref{appx:ablation-study-full}, without the resource-oriented specialist group, the search process may fail to recover from invalid parameter combinations that are close to optimal, such as the case of ``atax-medium" and ``syr2k". Having two distinct groups of specialists result in more diverse parameter update proposals at each search iteration.

We observe that the performance drop is not significant when one group of specialists is removed. This is because our router is still functioning by selecting candidate tasks for performance-oriented specialists. Our empirical results on replacing LLM-based agents with simple heuristics (\textsection{} \ref{exp:abl_llm_agent}) and direct generation (\textsection{} \ref{exp:generation}) further strengthen the advantage of having the router-specialists interaction.

\subsubsection{Ablating Necessity of LLM-based Agents}\label{exp:abl_llm_agent}
\input{tables/ablation}

Following the same experimental setup, we report the latency of the configurations discovered by different combinations of ``active" LLM-based agents. Table~\ref{tab:results-ablation-study} presents the results, where we replace our ablated components with fixed, programmed heuristics. We then progressively add LLM-based agents to evaluate their individual contributions to the overall exploration process.

We observe that as each component is progressively added, the number of winning cases over the baseline method steadily increases. The most significant improvement comes from the specialists, which enhances performance by dynamically adjusting parallelism based on the exploration history.

Another substantial improvement is introduced by the critic. The critic’s ability to evaluate configuration similarity and provide exploration guidance to the router enables the framework to escape local optima. This leads to configurations that achieve up to $5\times$ better performance on ``jacobi-2d" compared to the A+S+R baseline.

Overall, implementing all components as LLM-based agents results in state-of-the-art performance, with a geometric mean speedup of $1.87\times$ and a winning ratio of $8/10$ against the baseline.

\subsubsection{Ablation Study Summary} 
Our ablation study shows that: (1) Each interaction between agents is necessary to achieve the best performance. (2) Replacing fixed, programmed heuristics with domain-knowledge + feedback-guided LLMs boosts the adaptivity of LLM-DSE to different workloads. In the next section, we show that LLM-based agents can also adapt to different backend toolchains efficiently.

\subsection{Generalizing to Other Toolchains}\label{exp:multi_bknd}
In Appendix, Table~\ref{tab:asic_vitis_summary}, we evaluate LLM-DSE with two other toolchains targeting ASIC or on-chip FPGA modules. To support the ASIC flow, we add two specialists that handles ``Array Type'' and ``Array Partition'', and update the DSE knowledge presented to the arbitrator. Our change in the prompts is minimal, as detailed in Appx. \ref{appx:multi_bknd}. LLM-DSE achieves substantial speedups compared with the default parameter optimization provided by the toolchains.

\subsection{Direct Generation with Zero-Shot and One-Shot Prompting}\label{exp:generation}
We compare our framework with the parameters directly generated by the language models from several different prompts. For each setup, we repeat the experiment $8$ times to mitigate the randomness. The results, presented in Appendix, Table~\ref{tab:results-llm-dse}, indicate that our framework significantly outperforms direct generation. We envision that, by leveraging the interaction experience accumulated through our framework, it will be possible to fine-tune LLMs for more effective direct configuration generation or to enhance exploration efficiency.

\subsection{Evaluating Token consumption}
Running LLM-DSE for 8 hours on the Merlin backend consumes approximately $4\times 10^5$ to $2\times 10^6$ input tokens and $4\times 10^3$ to $3\times 10^4$ output tokens across the 10 programs we evaluated. The cost for running GPT-4o is from $1$ USD to $7$ USD. A detailed breakdown is shown in Appendix \ref{appx:token}.

To further reduce token consumption, possible approaches include: (1) merging multiple routers' call into a single one, and/or (2) modifying the history curator to limit the number of candidate tasks presented at each iteration.

\subsection{Scaling to Larger Programs}

\begin{wraptable}{r}{0.45\textwidth}  
  \vspace{-1.5em}                     
  \centering
  \small
  \caption{Performance comparison on the Rosetta benchmark.}\label{tab:rosetta}
  \setlength{\tabcolsep}{6pt}
  \begin{tabular}{l r@{\hspace{1em}}r@{\hspace{1em}}r}
    \toprule
     & \textbf{A8} & \textbf{LLM-DSE} & \textbf{Speedup} \\
    \midrule
    conv2d       & 1.6e8    & 1.4e8 $\pm$ 4e6  & \textcolor{darkgreen}{1.13} \\
    spam-filter  & 7.7e6    & 3.6e6 $\pm$ 1e6  & \textcolor{darkgreen}{2.12} \\
    knn          & 1.3e8    & 1.1e8 $\pm$ 4e4  & \textcolor{darkgreen}{1.15} \\
    3d-rendering & 9.9e6    & 1.2e7 $\pm$ 4e6  & 0.81 \\
    \midrule
    \textbf{Geomean} 
                 & 3.6e7    & 2.9e7           & \textcolor{darkgreen}{1.22} \\
    \bottomrule
  \end{tabular}
  \vspace{-1em}
\end{wraptable}
Despite being a publicly released dataset, kernels in HLSyn are small compared to real-world designs. To study the scalability of our approach, we also tested on larger programs from the Rosetta benchmark~\citep{rosetta}. The Conv2D, Spam Filter, KNN, and 3D Rendering programs contain 118, 126, 178, and 304 lines of code (LoC), respectively, compared to only 77 LoC in the largest HLSyn benchmark we evaluated. The Conv2D benchmark features up to seven levels of nested loops, whereas HLSyn programs contain at most three levels of nesting.

\begin{wraptable}{r}{0.45\textwidth}  
  \centering
  \small
  \caption{Comparison against RALAD.}\label{tab:ralad}
  \setlength{\tabcolsep}{4pt}
  \begin{tabular}{lccc}
    \toprule
    \textbf{Speedup} & \textbf{RALAD} & \textbf{A8} & \textbf{LLM-DSE} \\
    \midrule
    2mm          & 1.00       & 3.65    & 37.50  \\
    3mm          & timeout    & –       & –      \\
    adi          & 1.00       & 54.39   & 12.47  \\
    atax         & 1.00       & 412.91  & 412.91 \\
    bicg         & 1.00       & 72.84   & 72.84  \\
    correlation  & 1.00       & 112.81  & 128.82 \\
    \midrule
    \textbf{Average}  & –      & 131.32  & 132.91 \\
    \textbf{Geomean}  & –      & 58.32   & 71.06  \\
    \bottomrule
  \end{tabular}
  \vspace{-1em}
\end{wraptable}

Table \ref{tab:rosetta} presents the result. LLM-DSE outperforms heuristic driven approach by a geometric mean of $1.22\times$. While handling large dimensions is a challenge for most search algorithms, we have made several efforts to mitigate this issue. (1) The router is equipped with a “history curator” which selects and presents only K candidate designs per iteration, preventing context length explosion. (2) Each specialist focuses on a single pragma within one loop. (3) The critic compares only design pairs that differ by a single parameter. These efforts help us scale to larger programs with better performance.

\subsection{Comparison with Existing LLM-based Methods}
We compare against RALAD~\citep{ralad}, a retrieval-augmented method that integrates domain knowledge from relevant textbooks. 
While the database of the RALAD framework is not open-sourced, we take the open-source configuration generated by their framework and compare it against LLM-DSE. We also include the AutoDSE-8 baseline for reference. 
As shown in Table~\ref{tab:ralad}, LLM-DSE outperforms RALAD by a large margin, validating our observation that direct design generation usually performs poorly in this task.

\section{Related Work}
\subsection{Large Language Model Agents for Chip Design}
There has been growing interest in leveraging LLMs for chip design. Various techniques~\citep{chatgpt, cot, react, reflexion, comm, metagpt, camel, zeroshotcot} have been specialized or improved to tackle challenges across the design flow, from front-end development to back-end implementation. LaMAGIC~\citep{chang2024lamagic} and AnalogCoder~\citep{lai2024analogcoder} focus on generating analog circuits, while several other studies explore the generation of hardware description languages (HDL) such as Verilog~\citep{verilogeval, thakur2024verigen, zhang2024mg, cui2024origen, verilogcoder}. Our method leverages HLS tools to generate HDL in a Correct-by-Construction manner. In the context of HLS, C2HLSC~\citep{collini2024c2hlsc} and HLS-Repair~\citep{hlsrepair} investigate the transformation of arbitrary C/C++ code into valid HLS-C code, while HLSPilot~\citep{hlspilot} focuses on using LLMs to separate the software and hardware regions. To the best of our knowledge, LLM-DSE is the first multi-agent framework dedicated to optimizing the performance of hardware designs using HLS.

\subsection{Automatic Optimization of HLS Directives}
AutoDSE~\citep{autodse} leverages domain expertise to address the challenge of HLS design space exploration. More recently, model-based approaches have gained attention. The HLSyn benchmark~\citep{hlsyn} is a comprehensive dataset for applying machine learning to HLS optimization. Various models have been proposed to model HLS designs~\citep{ustun2020accurate, wu2022high, wu2022ironman, sohrabizadeh2022automated, sohrabizadeh2023robust, murphy2024balor}. Among them, Balor~\citep{murphy2024balor} achieved state-of-the-art results, securing first place in the ML4HLS contest~\citep{hlscontest}. However, Balor does not incorporate an exploration component for design optimization, making a direct comparison infeasible. HARP enhances graph-based HLS modeling through a hierarchical structure~\citep{sohrabizadeh2023robust}. ProgSG and CompareXplore~\citep{qin2024cross, bai2024learning} further improve HLS modeling by incorporating multi-modality and design ranking techniques. Hier-MoE~\citep{li2024hierarchical} investigates the domain generalization problem. In terms of design optimization, ~\citep{sun2022correlated} apply a standard Bayesian optimization framework, and Ironman~\citep{wu2022ironman} employs reinforcement learning for resource assignment. It does not consider the performance optimization directives explored in this work. Active-CEM~\citep{active-cem} introduces selective evaluation in the Cross-Entropy Method (CEM), while other approaches leverage mathematical models~\citep{pouget2024automatic, basalama2025stream, ye2024hida} to optimize designs. These models could serve as powerful external tools to further enhance the design efficiency of our framework.

\section{Limitation}\label{sec:limit}

LLM-DSE primarily focuses on optimizing the parameters of DSAs. One limitation is that it currently does not consider code transformations, which constitute a larger search space. Addressing this broader search scope is left to future work.

Another limitation is that, despite significantly enhancing the efficiency of the parameter search process, LLM-DSE still requires several hours to obtain good performance. Future research may focus on further improving the sample efficiency.

\section{Conclusion}\label{sec:conclude}

In this paper, we propose LLM-DSE, a new paradigm for designing domain specific hardware accelerators through efficient navigation in the design space guided by collaborative agents. 


This research does not introduce any new societal impacts, including those related to fairness, ethics, or privacy, none of which we believe require explicit discussion here.

\newpage
\bibliographystyle{plainnat}
\bibliography{llm-dse}

\newpage
\appendix
\section{HLS and directive parameter optimization}\label{appx:hls}

This is a brief introduction of the HLS design tools with directive parameters.

\begin{figure*}[h]
    \centering
    \includegraphics[width=0.8\textwidth]{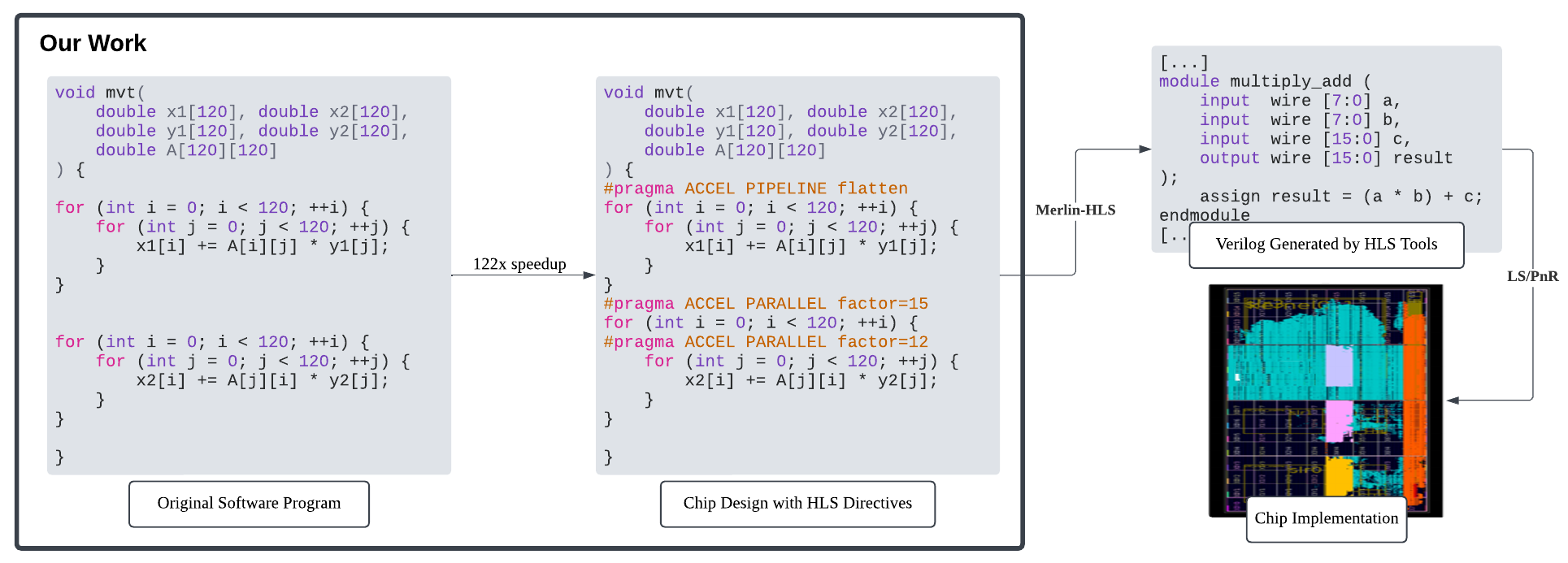}
    \caption{The Merlin Compiler simplifies hardware development by allowing designers to specify hardware directive parameters (``\#pragma ACCEL") within a given template. Still, due to the interdependent nature of hardware directives and the vast combinatorial space of possible parameters, identifying an efficient parameter set remains a challenging task, even for experienced designers.}
    \label{fig:merlin_pragma}
\end{figure*}

Figure~\ref{fig:merlin_pragma} illustrates a chip prototyping process on FPGAs and highlights the scope of our work. We utilize the Merlin Compiler, which processes a software program annotated with Merlin-HLS directives (parameters). The compiler then produces hardware description language (HDL) implementations, such as Verilog. These HDLs subsequently undergo further processing to be fully realized on-chip. Our work focuses on the initial stage of the chip design process, where the objective is to identify the optimal combination of HLS parameters that lead to high-performance DSA implementations.

Table \ref{tab:design_space_comparison} lists three different toolchains we support. The ``Merlin'' toolchain is where most existing baselines are built upon. The ``Vitis'' toolchain supports FPGA devices, while the ``Stratus'' toolchain targets ASIC devices. Despite sharing important parameter types such as ``Parallel" and ``Pipeline", their implications are quite different. For example, in Merlin, ``pipeline=cg'' will automatically apply double-buffering to overlap off-chip memory access and computation. However, in Vitis, ``Pipeline'' indicates synthesizing on-chip pipelined module, and is handled automatically. Moreover, in Merlin and Vitis, two parameter types ``Array Partition'' and ``Array Type'' are handled automatically. However, in Stratus, these two types of parameters need to be tuned. Our framework supports all three toolchains with minimal specialization efforts (\textsection{} \ref{appx:multi_bknd}), demonstrating the generalizability of our approach. 

\begin{table}[h!]
\centering
\caption{\small Comparison of parameter search spaces between different toolchains. The Merlin Compiler~\citep{merlin_compiler} synthesize end-to-end DSAs with off-chip memory transfer, while Vitis HLS~\citep{vitis24} synthesize on-chip modules. The Stratus HLS~\citep{stratus} targets ASICs, while Merlin and Vitis HLS targets FPGAs.}
\begin{tabular}{lccc}
\toprule
\textbf{Parameter Type} & \textbf{Merlin} & \textbf{Vitis} & \textbf{Stratus} \\
\midrule
Parallel & factor=int & factor=int & factor=int \\
Pipeline & off/fg/cg & auto & off/hs/ss \\
Tile     & yes       & no            & no         \\
Array Partition & auto      & auto          & separate/off \\
Array Type      & auto      & auto          & mem/reg     \\
\midrule
\textbf{Summary}  & \makecell[c]{\small FPGA: End-to-end with\\ off-chip memory transfer}
         & \small FPGA: On-chip
         & \small ASIC: On-chip\\
\bottomrule
\end{tabular}
\label{tab:design_space_comparison}
\end{table}

\begin{figure*}[h]
\centering
\includegraphics[width=0.95\columnwidth]{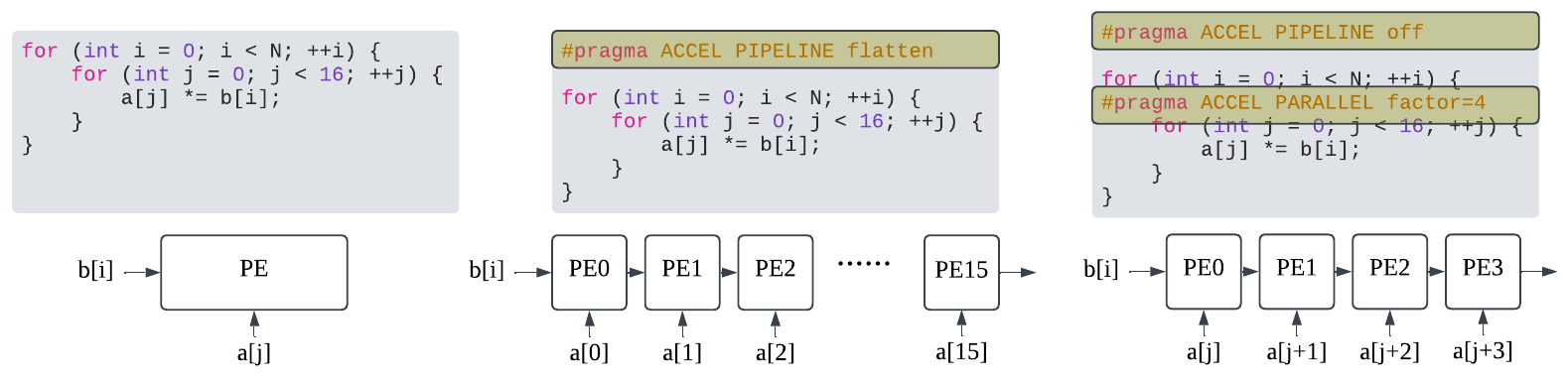}
\caption{\small Different combination of HLS parameters means different micro-architecture design. Here, each processing element (PE) multiply $a[j]$ with $b[i]$. When different directive parameters are selected, the underlying hardware will also be different, resulting in different latency and resource consumption.}
\label{fig:pragma_to_arch}
\end{figure*}

Figure~\ref{fig:pragma_to_arch} illustrate how different directives will result in different hardware architectures when using the Merlin Compiler as the backend toolchain. Note that the ``\#pragma ACCEL PIPELINE flatten" in the middle figure is equivalent to ``PIPELINE mode=fg" in Merlin. It is also equivalent to applying ``PIPELINE mode=cg" to loop $i$, and ``PARALLEL factor=16" to loop $j$. This results in the generation of $16$ processing elements (PEs) that process $a[i]$ parallelly. In contrast, on the right-most figure, applying ``PARALLEL factor=4" to the inner loop $j$ will result in the generation of $4$ parallel PEs.

\section{Detailed Prompt Designs}
We provide details of the prompts designed for each agent. All the texts without the [] are examples of our raw inputs to the agents.  

\subsection{Router}
\begin{figure}[h]
\centering
\includegraphics[width=0.9\columnwidth]{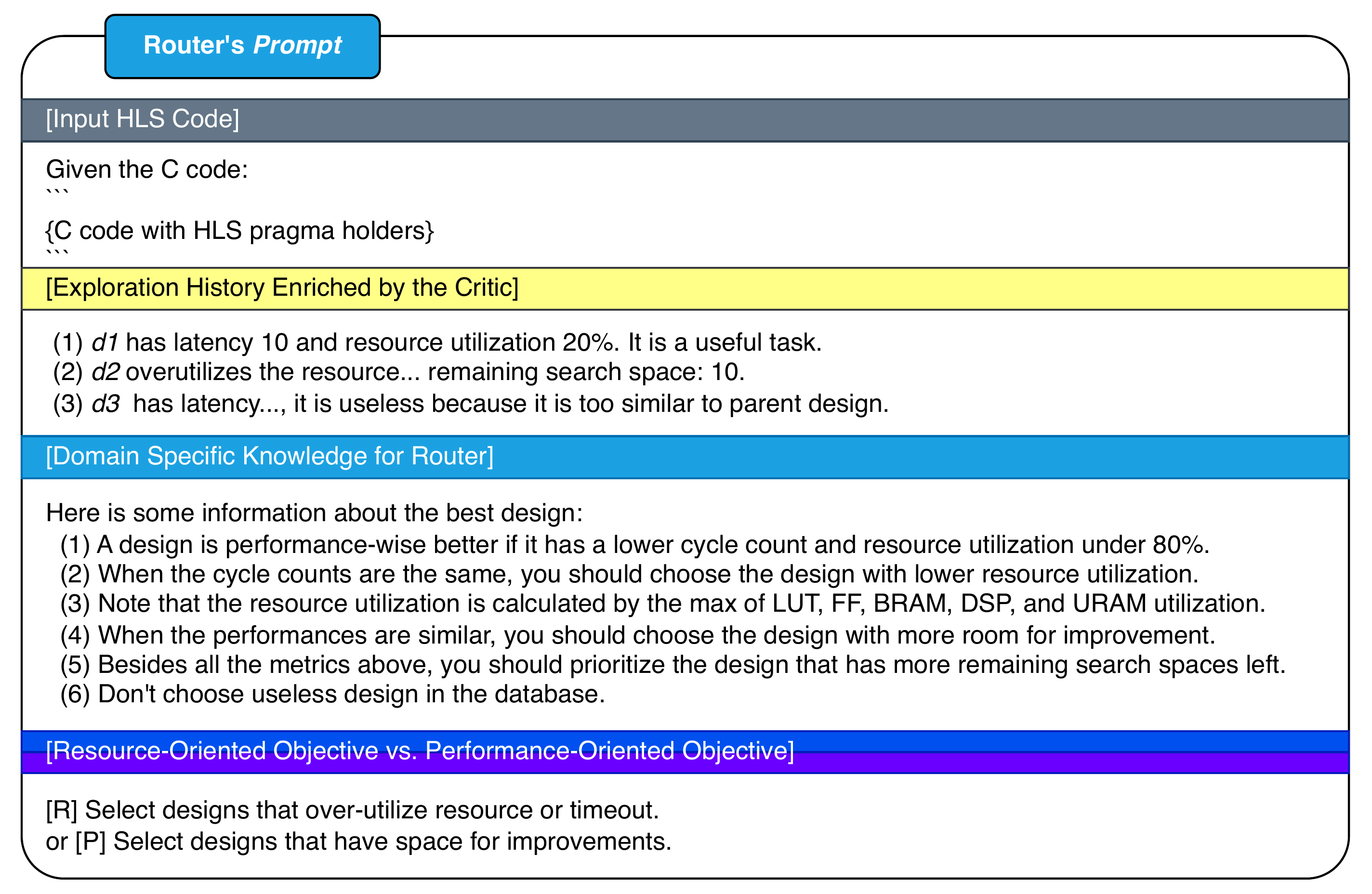}
\vspace{-5pt}
\caption{Router's prompt.}
\label{fig:router}
\end{figure}

The router is designed to select candidate tasks from the exploration history. As shown in~\ref{fig:router}, the router's prompt will take the input HLS code and the domain specific knowledge designed for the router as static information. Additionally, it will view the iteratively accumulated exploration history enriched by the critic's feedback. Finally, it will derive to different objectives based on the orientation.

\subsection{Specialists}

\begin{figure}[h]
\centering
\includegraphics[width=0.9\columnwidth]{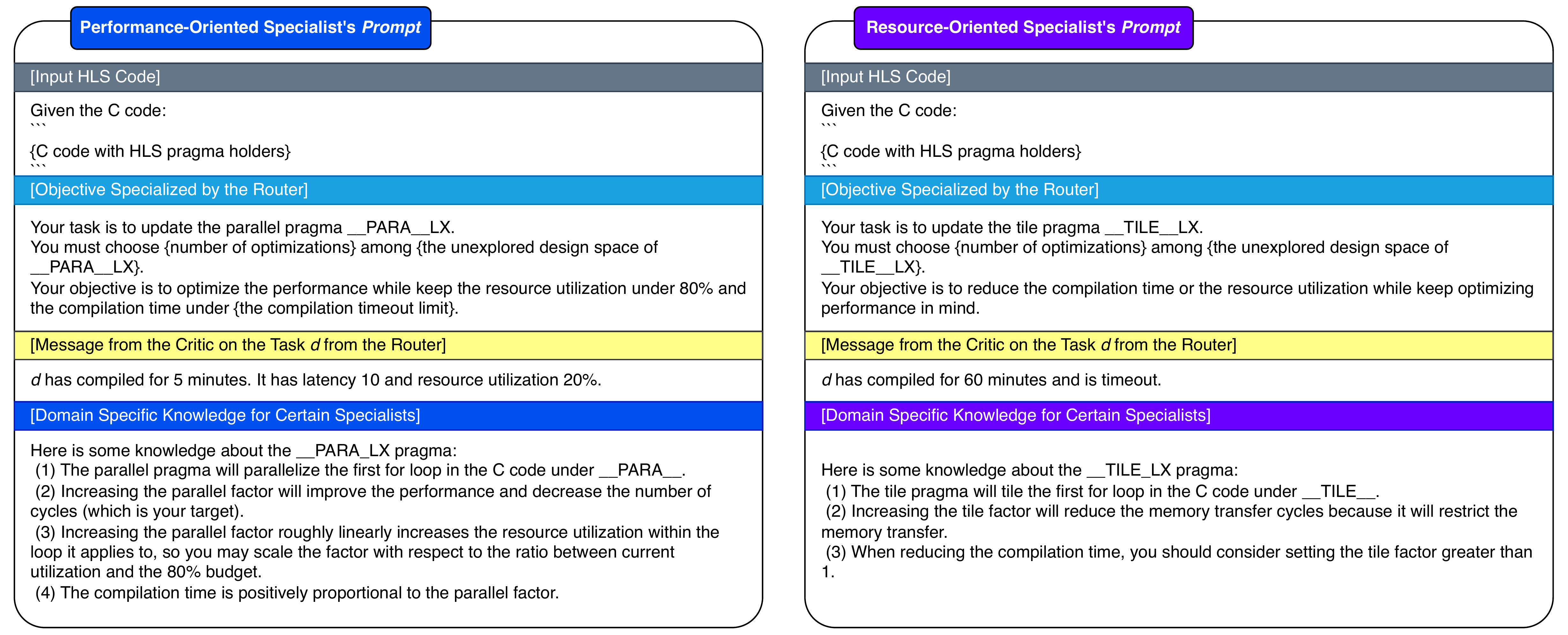}
\vspace{-5pt}
\caption{Specialists' prompt.}
\label{fig:specialists}
\end{figure}

For each specialist, we provide it with the message from the critic, the assigned task from the router, the exploration history and the domain specific knowledge of each different parameter type assigned to that specialist. The details are presented in Figure~\ref{fig:specialists}.

\subsection{Arbitrator}

\begin{figure}[h]
\centering
\includegraphics[width=0.95\columnwidth]{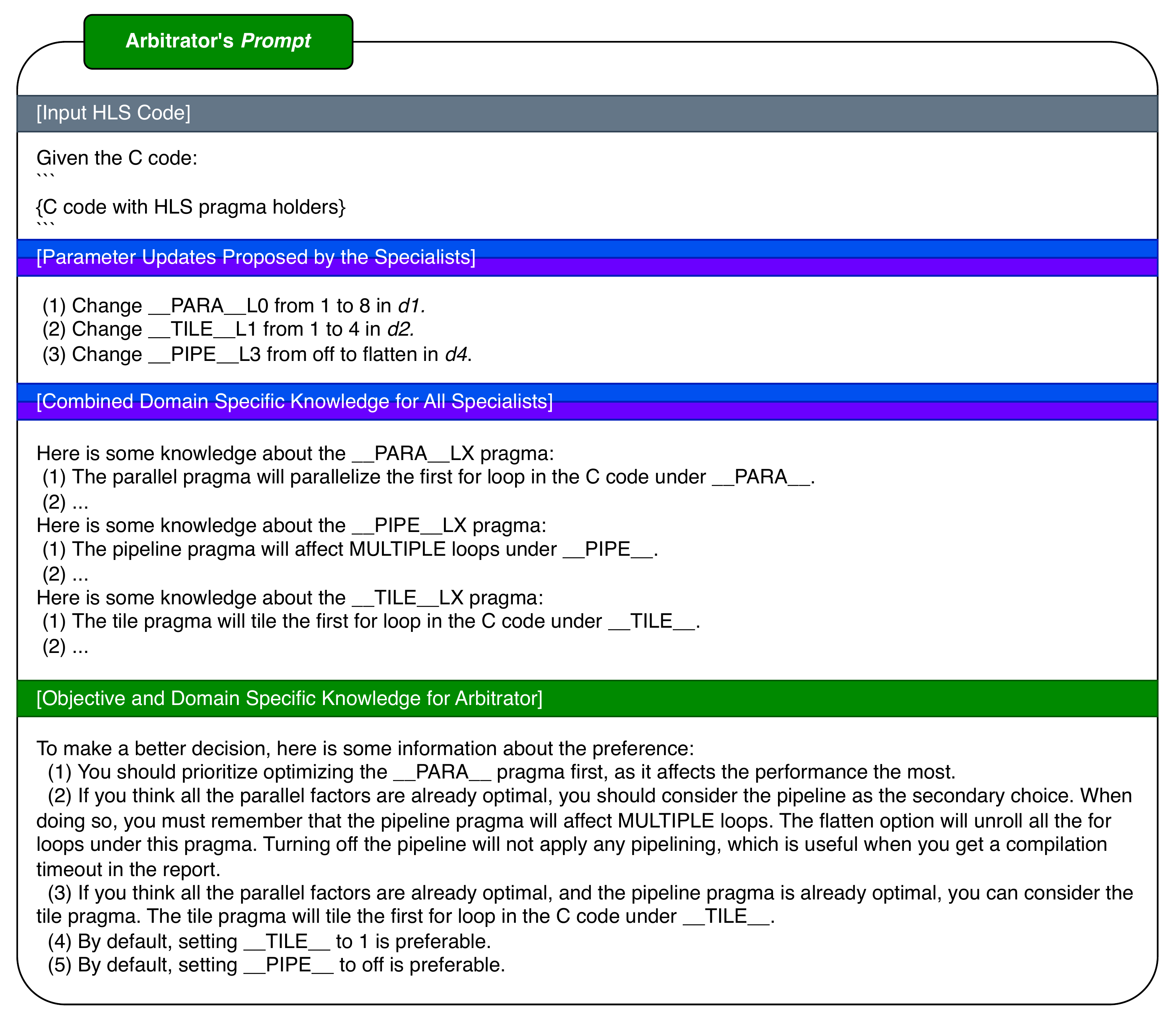}
\vspace{-5pt}
\caption{Arbitrator's prompt.}
\label{fig:arbitrator}
\end{figure}

As shown in Figure~\ref{fig:arbitrator}, the arbitrator is prompted with all the updates proposed by the specialists. Then, we present all the domain specific knowledge of each parameter type. In the end, we prompt the arbitrator with domain-specific heuristics to guide its decision making. 

\subsection{Prompt \& Code updates when supporting Stratus and Vitis}\label{appx:multi_bknd}

We demonstrate that with small modifications in the prompts, LLM-DSE could support other toolchains.

\subsubsection{From Merlin to Stratus}

\begin{figure}[h]
\centering
\includegraphics[width=0.9\columnwidth]{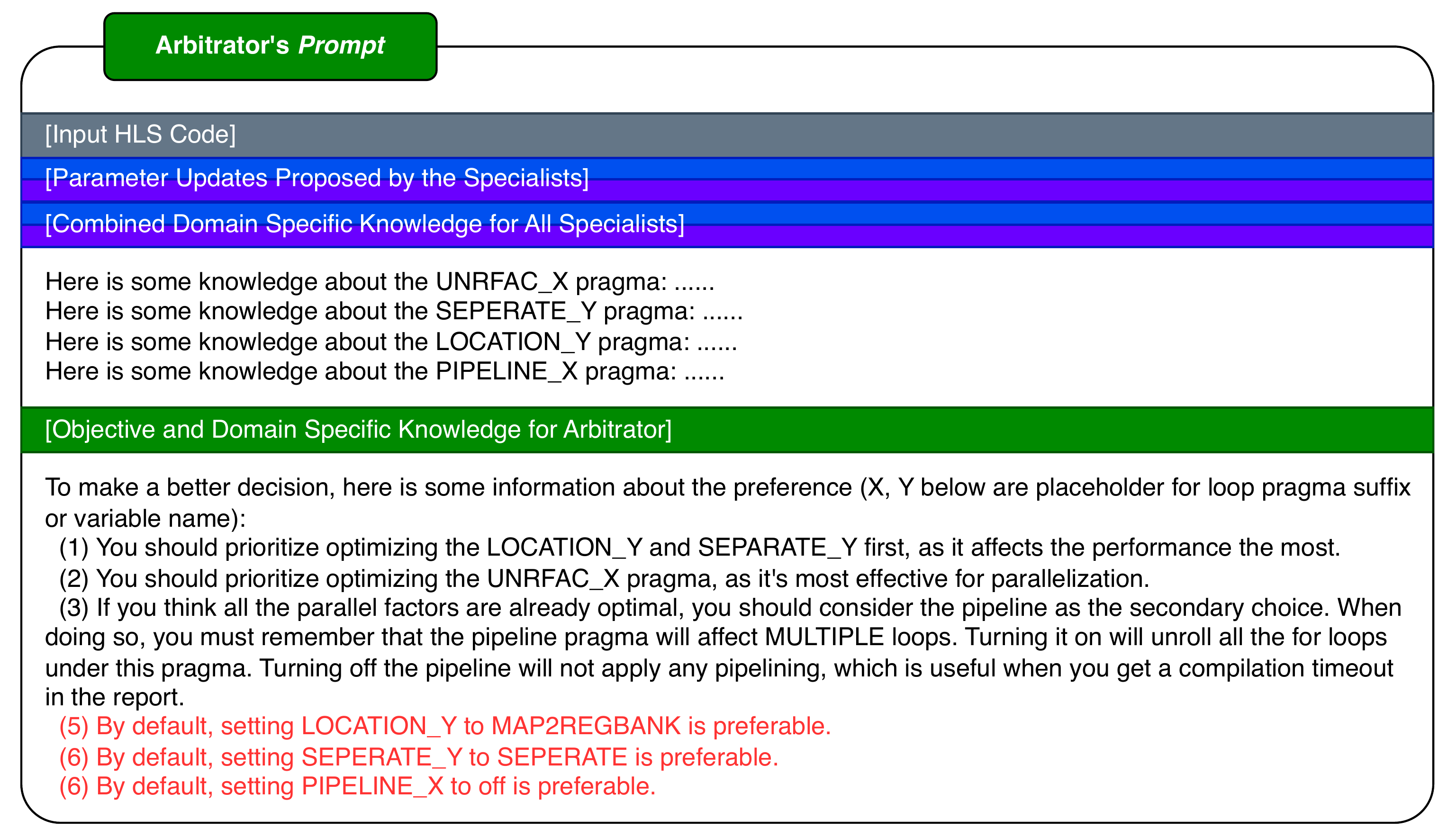}
\vspace{-5pt}
\caption{Stratus Arbitrator's prompt.}
\label{fig:stratus arbitrator}
\end{figure}

When adapting LLM-DSE to Stratus, we add two specialists to handle the ``Array Parition" and ``Array Type" parameters. Their prompts are included in our open-source code. We also update the heuristics provided to the arbitrator, as shown in Figure~\ref{fig:stratus arbitrator}.

\subsubsection{From Merlin to Vitis}

\begin{figure}[h]
\centering
\includegraphics[width=0.9\columnwidth]{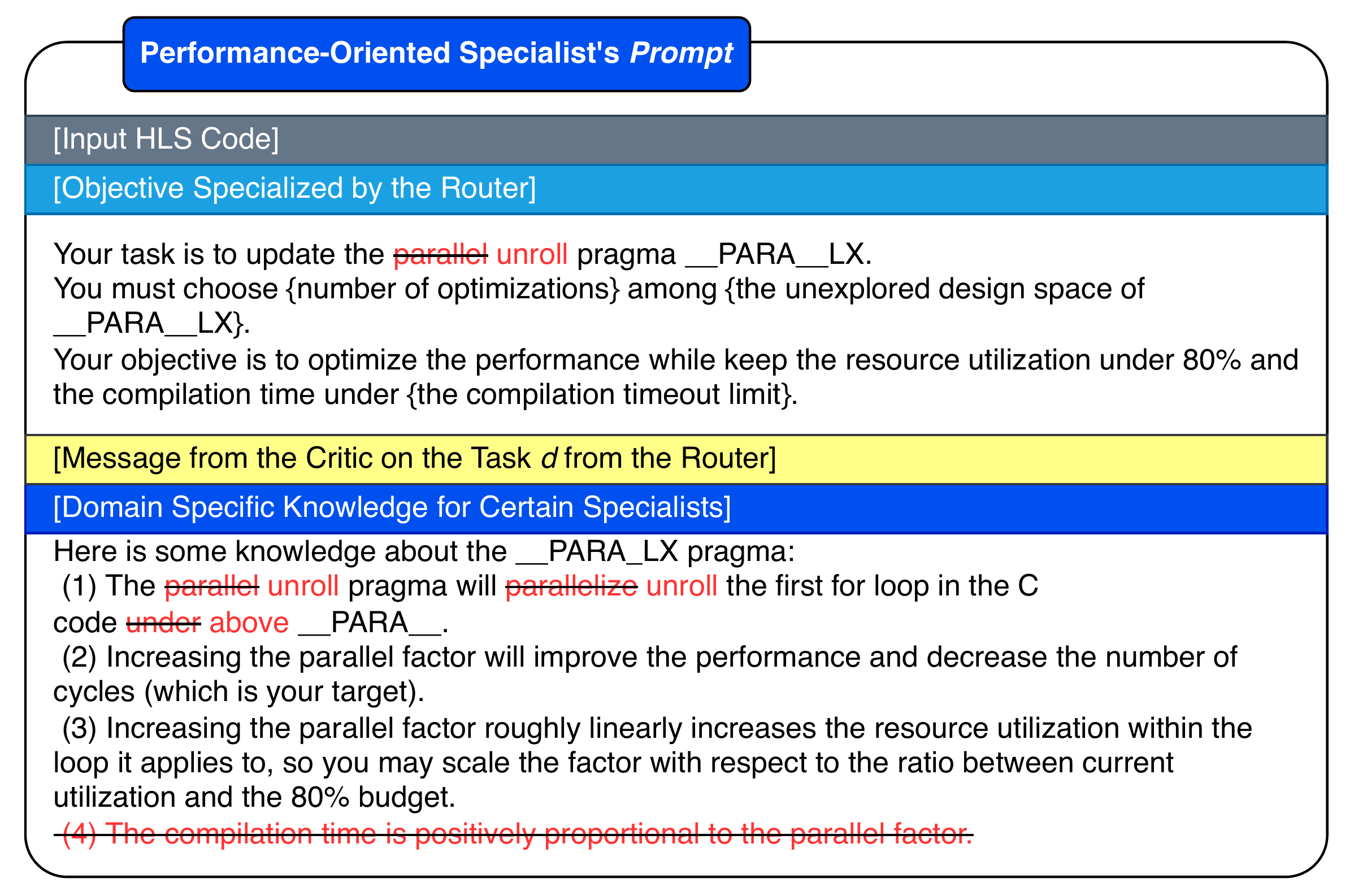}
\vspace{-5pt}
\caption{Vitis Specialist's prompt.}
\label{fig:vitis specialist}
\end{figure}

When supporting the Vitis toolchain, we only need to modify the prompt of the ``PARALLEL" specialist. The detailed modifications are illustrated in Figure~\ref{fig:vitis specialist}.

\section{Detailed Experiment Results}

\subsection{Ablating agent interactions}
\label{appx:ablation-study-full}

We list detailed result breakdown in the ablation study on agent interactions.

\input{tables/ablation-arch}

\subsection{Generalizing to other toolchains}

Table~\ref{tab:asic_vitis_summary} below presents the detailed results when adapting LLM-DSE to Stratus and Vitis.

\begin{table}[h!]
\centering
\caption{\small Performance on ASIC (Stratus) and Vitis flow. The raw latency and the speedup is reported.}
\begin{tabular}{lrrrrrr}
\toprule
\multirow{2}{*}{\textbf{Benchmark}} & \multicolumn{3}{c}{\textbf{Stratus}} & \multicolumn{3}{c}{\textbf{Vitis}} \\
\cmidrule(lr){2-4}\cmidrule(lr){5-7}
 & Default & LLM-DSE & Speedup & Default & LLM-DSE & Speedup \\
\midrule
3mm        & 611505   & 598839   & \textcolor{darkgreen}{1.02} & 1691966 & 908597  & \textcolor{darkgreen}{1.86}  \\
atax-med   & --       & --       &                          -- & 1290929 & 259023  & \textcolor{darkgreen}{4.98}  \\
covariance & 576002   & 569281   & \textcolor{darkgreen}{1.01} & 3915663 & 130525  & \textcolor{darkgreen}{30.00} \\
fdtd-2d    & 2285960  & 1713990  & \textcolor{darkgreen}{1.33} & 581010  & 225786  & \textcolor{darkgreen}{2.57}  \\
gemm-p     & 714405   & 359863   & \textcolor{darkgreen}{1.99} & 352290  & 66640   & \textcolor{darkgreen}{5.29}  \\
gemver-med & --       & --       &                          -- & 2400477 & 444558  & \textcolor{darkgreen}{5.40}  \\
jacobi-2d  & 3130040  & 2506475  & \textcolor{darkgreen}{1.42} & 638705  & 216307  & \textcolor{darkgreen}{2.95}  \\
symm-opt   & 631202   & 621661   & \textcolor{darkgreen}{1.02} & --      & 1916576 & \textcolor{darkgreen}{--}    \\
syr2k      & 1969600  & 1568840  & \textcolor{darkgreen}{1.26} & 406490  & 53653   & \textcolor{darkgreen}{7.58}  \\
trmm-opt   & 314461   & 301379   & \textcolor{darkgreen}{1.04} & 33984   & 33980   & \textcolor{darkgreen}{1.00}  \\
\midrule
\textbf{Geomean}    & 957056   & 792066 (17156)   & \textcolor{darkgreen}{1.23} &   & 194992 (512) & \textcolor{darkgreen}{4.26}  \\
\textbf{Win \%}     &          &          & 100\% &   &  & 100\%        \\
\bottomrule
\end{tabular}
\label{tab:asic_vitis_summary}
\end{table}

\subsection{Zero-shot and one-shot experiment}

Table~\ref{tab:results-llm-dse} shows the results of our zero-shot and one-shot experiments.

\input{tables/direct-generation}

\section{Token Consumption of LLM-DSE}
\label{appx:token}

\subsection{Number of input/output tokens and overhead over necessary information}

\begin{table}[h!]
\centering
\vspace{-5pt}
\caption{Token usage breakdown and overhead over baseline information.}
\begin{tabular}{lrrrrrr}
\toprule
\textbf{Benchmark} & \textbf{\makecell[l]{Input\\ Total}} & \textbf{\makecell[l]{Output\\ Total}} & \textbf{\makecell[l]{Baseline:\\ Code+Log}} & \textbf{\makecell[l]{Code\\ Input}} & \textbf{\makecell[l]{Log\\ Input}} & \textbf{\makecell[l]{Delta to\\ Baseline}} \\
\midrule
3mm & 2633467 & 27500 & 2049692 & 591318 & 1458374 & 1.28 \\
atax-med & 193652 & 4189 & 121612 & 30438 & 91174 & 1.59 \\
covariance & 1111239 & 15346 & 821864 & 241500 & 580364 & 1.35 \\
fdtd-2d & 1103020 & 11196 & 860876 & 230480 & 630396 & 1.28 \\
gemm-p & 478037 & 9058 & 325400 & 92000 & 233400 & 1.46 \\
gemver-med & 861753 & 12716 & 627435 & 187425 & 440010 & 1.37 \\
jacobi-2d & 696858 & 10513 & 498133 & 145866 & 352267 & 1.40 \\
symm-opt & 719798 & 11513 & 501110 & 148785 & 352325 & 1.44 \\
syr2k & 400229 & 8150 & 268876 & 84298 & 184578 & 1.49 \\
trmm-opt & 436723 & 7789 & 287416 & 58426 & 228990 & 1.52 \\
3d-rendering & 2689207 & 14677 & 2347689 & 1632015 & 715674 & 1.14 \\
knn & 3293064 & 27511 & 2708889 & 1418508 & 1290381 & 1.21 \\
spam-filter & 1871112 & 21665 & 1439902 & 580932 & 858970 & 1.29 \\
conv2d & 4721311 & 27877 & 3865126 & 1328925 & 2536201 & 1.22 \\
\bottomrule
\end{tabular}
\end{table}

Baseline represents the number of tokens consumed by necessary information, including input code and compiler logs. Results show that our approach has acceptable overhead over the baseline, indicating that our knowledge database and instructions only introduce slight overhead over the necessary information. In addition, we observe that the number of input tokens we consume is much more than the output tokens. This is because we need to put the code and the compiler logs in the input prompt for the agent to make decisions.

\subsection{Breakdown Among Agents}

\begin{table}[h!]
\centering
\caption{Token usage ratio by agent.}
\begin{tabular}{lrrrr}
\toprule
\textbf{Benchmark} & \textbf{router} & \textbf{specialists} & \textbf{critic} & \textbf{arbitrator} \\
\midrule
3mm & 0.25 & 0.60 & 0.03 & 0.11 \\
atax-med & 0.24 & 0.53 & 0.13 & 0.10 \\
covariance & 0.28 & 0.57 & 0.05 & 0.10 \\
fdtd-2d & 0.22 & 0.64 & 0.04 & 0.10 \\
gemm-p & 0.27 & 0.55 & 0.08 & 0.10 \\
gemver-med & 0.29 & 0.55 & 0.05 & 0.11 \\
jacobi-2d & 0.25 & 0.58 & 0.06 & 0.11 \\
symm-opt & 0.26 & 0.56 & 0.08 & 0.09 \\
syr2k & 0.29 & 0.52 & 0.08 & 0.10 \\
trmm-opt & 0.22 & 0.58 & 0.10 & 0.10 \\
3d-rendering & 0.42 & 0.49 & 0.02 & 0.07 \\
knn & 0.35 & 0.54 & 0.03 & 0.08 \\
spam-filter & 0.30 & 0.56 & 0.05 & 0.09 \\
conv2d & 0.27 & 0.61 & 0.02 & 0.11 \\
\bottomrule
\end{tabular}
\end{table}

We observe that the ``router" consumes the most tokens. This is because (1) at each iteration, we instantiate the router multiple times for each specialist, with different instructions (2) the router requires multiple compiler logs to be presented to it to make selection. We would like to note that we have already made a preliminary effort to reduce the number of tokens consumed. Specifically, we parse the logs and keep only the necessary information (the design configuration, the latency and resource consumption).

The ``specialists" also consumes a lot of tokens. This is because we have different specialists for each type of parameter and for each loop.

\subsection{Number of tokens per iteration}

\begin{table}[h!]
\centering
\caption{Token usage per iteration.}\label{tab:tok_iter}
\begin{tabular}{lrrrr}
\toprule
\textbf{Benchmark} & \textbf{Iter 4} & \textbf{Iter 6} & \textbf{Iter 8} & \textbf{Iter 10} \\
\midrule
3mm & 161013 & 171271 & 188029 & 188235 \\
atax-med & 24223 & 25065 & 22620 & - \\
covariance & 81104 & 81339 & 92382 & 93948 \\
fdtd-2d & 122421 & 130643 & 133686 & 139123 \\
gemm-p & 40817 & 44181 & 45651 & 42814 \\
gemver-med & 83235 & 86862 & 94297 & 90495 \\
jacobi-2d & 64762 & 66134 & 65679 & 64248 \\
symm-opt & 42728 & 42871 & 40986 & 40481 \\
syr2k & 46155 & 46408 & 45033 & - \\
trmm-opt & 31957 & 31150 & 35429 & 36000 \\
3d-rendering & 191473 & 225502 & 210687 & 206317 \\
knn & 199441 & 194102 & 188894 & 182003 \\
spam-filter & 92556 & 94518 & 88773 & 90231 \\
conv2d & 319065 & 404600 & 439850 & 463640 \\
\bottomrule
\end{tabular}
\end{table}

As shown in Table \ref{tab:tok_iter}, LLM-DSE maintains a stable token consumption throughout the exploration process. This is because we implement a history curator for the router. This avoids the explosion of context length as more configurations are explored.

\section{Case study}
\label{appx:case}

We present two case studies of LLM-DSE’s advantage over search algorithms with a fixed heuristic. On the trmm-opt and the syr2k kernel, LLM-DSE could find a config with more than 2$\times$ speedup compared with AutoDSE, even when we run AutoDSE 3$\times$ longer than LLM-DSE.

We include below the best configs found by AutoDSE (after 24 hours, denoted as Auto24 in the paper) and LLM-DSE (after 8 hours), to demonstrate that LLM-DSE --- through its novel architecture that integrates domain knowledge into the search process --- can effectively overcome the limitations of fixed-heuristic-driven approaches.

\definecolor{codegray}{gray}{0.95}
\lstset{
  backgroundcolor=\color{codegray},
  basicstyle=\ttfamily\footnotesize,
  breaklines=true,
  frame=single,
  language=C,
  keywordstyle=\color{blue}\bfseries,
  commentstyle=\color{gray},
  showstringspaces=false
}

\subsubsection*{Program: syr2k}

\textit{Config found by AutoDSE}
\begin{lstlisting}
void kernel_syr2k(...) {
  int i, j, k;
#pragma ACCEL PIPELINE off
#pragma ACCEL TILE FACTOR=1
#pragma ACCEL PARALLEL FACTOR=1
  for (i = 0; ...) {
#pragma ACCEL PARALLEL FACTOR=8
    for (j = 0; ...) {
      //...
    }
#pragma ACCEL PIPELINE
#pragma ACCEL TILE FACTOR=1
#pragma ACCEL PARALLEL FACTOR=4
    for (k = 0; ...) {
#pragma ACCEL PARALLEL FACTOR=80
      for (j = 0; ...) {
        //...
      }
    }
  }
}
\end{lstlisting}

\textit{Config found by LLM-DSE}
\begin{lstlisting}
void kernel_syr2k(...) {
  int i, j, k;
#pragma ACCEL PIPELINE off
#pragma ACCEL TILE FACTOR=1
#pragma ACCEL PARALLEL FACTOR=1
  for (i = 0; ...) {
#pragma ACCEL PARALLEL FACTOR=20
    for (j = 0; ...) {
      //...
    }
#pragma ACCEL PIPELINE off
#pragma ACCEL TILE FACTOR=1
#pragma ACCEL PARALLEL FACTOR=10
    for (k = 0; ...) {
#pragma ACCEL PARALLEL FACTOR=40
      for (j = 0; ...) {
        //...
      }
    }
  }
}
\end{lstlisting}

For this program, AutoDSE prioritizes inner loop unrolling (factor=80), and stays in this local mode despite exploring for more time. LLM-DSE could balance the unrolling of inner and outer loops with the synergy between the specialists and the arbitrator, resulting in finding configs with much better performance in much shorter times.

\subsubsection*{Program: trmm-opt}

\textit{Config found by AutoDSE}
\begin{lstlisting}
void kernel_trmm(...)
{
#pragma ACCEL PIPELINE off
#pragma ACCEL TILE FACTOR=1
#pragma ACCEL PARALLEL FACTOR=1
  for (int i = 0; ...) {
#pragma ACCEL PIPELINE flatten
#pragma ACCEL TILE FACTOR=1
#pragma ACCEL PARALLEL FACTOR=1
    for (int j = 0; ...) {
      //...
#pragma ACCEL PARALLEL reduction=sum FACTOR=1
      for (int k = 0; ...) {
        //...
      }
      //...
    }
  }
}
\end{lstlisting}

\textit{Config found by LLM-DSE}
\begin{lstlisting}
void kernel_trmm(...)
{
#pragma ACCEL PIPELINE off
#pragma ACCEL TILE FACTOR=1
#pragma ACCEL PARALLEL FACTOR=15
  for (int i = 0; ...) {
#pragma ACCEL PIPELINE flatten
#pragma ACCEL TILE FACTOR=1
#pragma ACCEL PARALLEL FACTOR=8
    for (int j = 0; ...) {
      //...
#pragma ACCEL PARALLEL reduction=sum FACTOR=1
      for (int k = 0; ...) {
        //...
      }
      //...
    }
  }
}
\end{lstlisting}

In this case, LLM-DSE is able to optimize the value of the pragmas more extensively in each iteration, so it’s able to identify a configuration with both pipeline flattening and coarse-grained unrolling (factor=8). For AutoDSE, it will only optimize one directive at a time by moving to its adjacent value. It fails to explore the scheme of flattening and unrolling the same loop under the 24 hours' time budget, due to its programmed heuristic.

\end{document}

%% file: macro.tex
\usepackage{algorithm}
\usepackage{algorithmic}
\usepackage{amsmath}
\usepackage{amssymb}
\usepackage{dsfont}
\usepackage{graphicx}
\usepackage{soul}
\usepackage{listings}

\usepackage[dvipsnames]{xcolor}
\colorlet{darkgreen}{ForestGreen!90!black}

\usepackage{tikz}
\usepackage{subcaption}

\usepackage{threeparttable}

\usepackage{booktabs}
\usepackage{multicol}
\usepackage{multirow}
\usepackage{textcomp}

\usepackage{wrapfig}

%% file: tables/comparison.tex
\begin{table*}[b]
\centering
\small
\caption{Comparison of our LLM-DSE explorer against three baselines on the Merlin Compiler backend. Each experiment runs for 8 hours per program, is conducted twice, and we report the mean \(\pm\) std of the best configuration. S-X denotes speedup over baseline X.}
\label{tab:performance-evaluation-reordered}
\begin{tabular}{l
                r@{\,\,}l
                r@{\hspace{1em}}r
                r@{\hspace{1em}}r
                r@{\hspace{1em}}r
                r@{\hspace{1em}}r}
\toprule
\multirow{2}{*}{\textbf{Benchmark}}
  & \multicolumn{2}{c}{\textbf{LLM-DSE}}
  & \multicolumn{2}{c}{\textbf{vs.\ A24}}
  & \multicolumn{2}{c}{\textbf{vs.\ A8}}
  & \multicolumn{2}{c}{\textbf{vs.\ H24}}
  & \multicolumn{2}{c}{\textbf{vs.\ H8}} \\
\cmidrule(lr){2-3}\cmidrule(lr){4-5}\cmidrule(lr){6-7}\cmidrule(lr){8-9}\cmidrule(lr){10-11}
  & Mean    & \(\pm\)Std
  & Time    & \(S\)
  & Time    & \(S\)
  & Time    & \(S\)
  & Time    & \(S\) \\
\midrule
3mm           & 26\,539  & \(\pm\)7\,351  
              & 128\,908 & \textcolor{darkgreen}{4.86}  
              & 189\,570 & \textcolor{darkgreen}{7.14}  
              &  9\,762  & 0.37  
              & 11\,083  & 0.42     \\
atax-med      &131\,118  & \(\pm\)60\,813 
              &  88\,117 & 0.67  
              & 232\,075 & \textcolor{darkgreen}{1.77}  
              &  92\,991 & 0.71  
              & 298\,358 & \textcolor{darkgreen}{2.28} \\
covariance    & 29\,344  & \(\pm\)0      
              &  22\,668 & 0.77  
              &  29\,668 & \textcolor{darkgreen}{1.01}  
              &  22\,168 & 0.76  
              &  31\,480 & \textcolor{darkgreen}{1.07} \\
fdtd-2d       & 12\,583  & \(\pm\)28     
              &  15\,603 & \textcolor{darkgreen}{1.24}  
              &  25\,054 & \textcolor{darkgreen}{1.99}  
              &  15\,603 & \textcolor{darkgreen}{1.24}  
              & 209\,123 & \textcolor{darkgreen}{16.62}\\
gemm-p        &  9\,120  & \(\pm\)98    
              &   9\,179 & \textcolor{darkgreen}{1.01}  
              &   9\,179 & \textcolor{darkgreen}{1.01}  
              &   9\,179 & \textcolor{darkgreen}{1.01}  
              &   9\,179 & 1.00     \\
gemver-med    &187\,339  & \(\pm\)30\,796
              & 148\,606 & 0.79  
              & 168\,086 & 0.90  
              & 265\,686 & \textcolor{darkgreen}{1.42}  
              &      –   & –        \\
jacobi-2d     &164\,284  & \(\pm\)0      
              & 164\,284 & 1.00  
              & 238\,164 & \textcolor{darkgreen}{1.45}  
              & 164\,284 & 1.00  
              & 206\,364 & \textcolor{darkgreen}{1.26} \\
symm-opt      & 13\,277  & \(\pm\)0      
              &  13\,277 & 1.00  
              &  13\,277 & 1.00  
              &  13\,277 & 1.00  
              &  15\,078 & \textcolor{darkgreen}{1.14} \\
syr2k         & 22\,739  & \(\pm\)5\,485 
              &  45\,501 & \textcolor{darkgreen}{2.00}  
              &  51\,581 & \textcolor{darkgreen}{2.27}  
              &  45\,501 & \textcolor{darkgreen}{2.00}  
              & 507\,348 & \textcolor{darkgreen}{22.31} \\
trmm-opt      &  3\,517  & \(\pm\)0      
              &   9\,387 & \textcolor{darkgreen}{2.67}  
              &  24\,567 & \textcolor{darkgreen}{6.99}  
              &   7\,395 & \textcolor{darkgreen}{2.10}  
              &  33\,032 & \textcolor{darkgreen}{9.39} \\
\midrule
\textbf{Average}
              &         &          
              &         & \textcolor{darkgreen}{1.60}  
              &         & \textcolor{darkgreen}{2.55}  
              &         & \textcolor{darkgreen}{1.16} 
              &         & \textcolor{darkgreen}{6.16} \\
\textbf{Geomean}
              &         &          
              &         & \textcolor{darkgreen}{1.30}  
              &         & \textcolor{darkgreen}{1.87}  
              &         & \textcolor{darkgreen}{1.04} 
              &         & \textcolor{darkgreen}{2.58} \\
\bottomrule
\end{tabular}
\end{table*}

%% file: tables/ablation.tex
\begin{table*}[b]
\centering
\small
\caption{\small Ablation: Necessity of LLM-based Agents. The baseline is the AutoDSE gradient‐based explorer. A: Arbitrator, S: Specialists, R: Router, C: Critic. For example, A+S means that A and S are implemented with LLM-based agents, while R and C are implemented with heuristics. All run under an 8-hour limit; we report average cycle counts and speedups \(S\) over the baseline.}\label{tab:results-ablation-study}
\begin{tabular}{l 
                r 
                r@{\hspace{1em}}r 
                r@{\hspace{1em}}r 
                r@{\hspace{1em}}r 
                r@{\hspace{1em}}r}
\toprule
\multirow{2}{*}{\textbf{Benchmark}}
  & \multirow{2}{*}{\textbf{Baseline(A8)}}
  & \multicolumn{2}{c}{\textbf{A}}
  & \multicolumn{2}{c}{\textbf{A+S}}
  & \multicolumn{2}{c}{\textbf{A+S+R}}
  & \multicolumn{2}{c}{\textbf{A+S+R+C}} \\
\cmidrule(lr){3-4}\cmidrule(lr){5-6}\cmidrule(lr){7-8}\cmidrule(lr){9-10}
  & 
  & \#Cycle & \(S\)
  & \#Cycle & \(S\)
  & \#Cycle & \(S\)
  & \#Cycle & \(S\) \\
\midrule
3mm            & 189\,570 & 357\,335           & 0.53      
                              &  14\,470           & \textcolor{darkgreen}{13.10}  
                              &  32\,094           & \textcolor{darkgreen}{5.91}   
                              &  26\,539           & \textcolor{darkgreen}{7.14}  \\
atax-medium    & 232\,075 &  88\,117           & \textcolor{darkgreen}{2.63}  
                              & 131\,124           & 1.77        
                              & 108\,327           & \textcolor{darkgreen}{2.14}   
                              & 131\,119           & 1.77      \\
covariance     &  29\,668 & 378\,893           & 0.08      
                              & 255\,778           & 0.12        
                              & 497\,104           & 0.06      
                              &  29\,344           & \textcolor{darkgreen}{1.01}  \\
fdtd-2d        &  25\,054 & 116\,303           & 0.22      
                              &  24\,963           & 1.00        
                              &  10\,751           & \textcolor{darkgreen}{2.33}   
                              &  12\,583           & \textcolor{darkgreen}{1.99}  \\
gemm-p         &   9\,179 &  20\,946           & 0.44      
                              &   8\,123           & \textcolor{darkgreen}{1.13}  
                              &   8\,827           & \textcolor{darkgreen}{1.04}  
                              &   9\,120           & \textcolor{darkgreen}{1.01}  \\
gemver-medium  & 168\,086 & 281\,207           & 0.60      
                              & 301\,756           & 0.56        
                              & 167\,416           & 1.00      
                              & 187\,339           & 0.90      \\
jacobi-2d      & 238\,164 & 394\,704           & 0.60      
                              & 625\,124           & 0.38        
                              & 625\,124           & 0.38      
                              & 164\,284           & \textcolor{darkgreen}{1.45}  \\
symm-opt       &  13\,277 &  13\,277           & 1.00      
                              & 185\,289           & 0.07        
                              &  13\,277           & 1.00      
                              &  13\,277           & 1.00      \\
syr2k          &  51\,581 &  45\,541           & \textcolor{darkgreen}{1.13}  
                              &  28\,699           & \textcolor{darkgreen}{1.80}  
                              &  29\,261           & \textcolor{darkgreen}{1.76}  
                              &  22\,740           & \textcolor{darkgreen}{2.27}  \\
trmm-opt       &  24\,567 &   4\,118           & \textcolor{darkgreen}{5.96}  
                              &   3\,517           & \textcolor{darkgreen}{6.99}  
                              &   3\,517           & \textcolor{darkgreen}{6.99}  
                              &   3\,517           & \textcolor{darkgreen}{6.99}  \\
\midrule
\textbf{Geomean}
               &          &                    & 0.69      
                              &                    & 0.95        
                              &                    & \textcolor{darkgreen}{1.24}  
                              &                    & \textcolor{darkgreen}{1.87}  \\
\textbf{Win \%}
               &          &                    & 30\%      
                              &                    & 50\%        
                              &                    & 60\%      
                              &                    & 80\%      \\
\bottomrule
\end{tabular}
\end{table*}

%% file: tables/ablation-arch.tex
\begin{table*}[hbtp]
\centering
\small
\caption{\small Ablation: Agent Interactions. We compare LLM-DSE to two simpler architecture: w/o arbitrator and only performance-oriented specialists. All run under an 8-hour limit; we report average cycle counts across two runs. \(S1\) denotes speedup over ``w/o arbitrator" and \(S2\) denotes speedup over ``only perf. specialists"}\label{tab:abl_agent_arch_full}
\begin{tabular}{lrrrrrr}
\toprule
\textbf{Benchmark} & \multicolumn{1}{c}{\textbf{w/o arbitrator}} & \multicolumn{1}{c}{\textbf{Only perf. specialists}} & \multicolumn{1}{c}{\textbf{LLM-DSE}} & \multicolumn{1}{c}{\textbf{\(S1\)}} & \multicolumn{1}{c}{\textbf{\(S2\)}} \\
\midrule
3mm & 388190 & 26433 & 26539 & 14.63 & 1.00 \\
atax-medium & 98222 & 174120 & 131119 & 0.75 & 1.33 \\
covariance & 29506 & 25824 & 29344 & 1.01 & 0.88 \\
fdtd-2d & 23863 & 16109 & 12583 & 1.90 & 1.28 \\
gemm-p & 10301 & 8997 & 9120 & 1.13 & 0.99 \\
gemver-medium & 179526 & 161315 & 187339 & 0.96 & 0.86 \\
jacobi-2d & 394704 & 185324 & 164284 & 2.40 & 1.13 \\
symm-opt & 17490 & 17490 & 13277 & 1.32 & 1.32 \\
syr2k & 30855 & 32101 & 22740 & 1.36 & 1.41 \\
trmm-opt & 3517 & 3517 & 3517 & 1.00 & 1.00 \\
\midrule
\textbf{Geomean} & 45399 (16159)      & 31675 (4410)      &  28728 (3311)     & 1.58 & 1.10 \\
\textbf{Win \%} &       &       &       & 70\% & 50\% \\
\textbf{Tie \%} &       &       &       & 10\% & 10\% \\
\bottomrule
\end{tabular}
\end{table*}

%% file: tables/direct-generation.tex
\begin{table*}[h]
\centering
\small
\caption{Comparison with Direct Generation}
\label{tab:results-llm-dse}
\begin{tabular}{l
                r
                rr
                rr
                rr}
\toprule
\multirow{2}{*}{\textbf{Benchmark}} &
\multicolumn{1}{c}{\textbf{DSE(A8)}} &
\multicolumn{2}{c}{\textbf{GPT-4o}} &
\multicolumn{2}{c}{\textbf{GPT-o1}} &
\multicolumn{2}{c}{\textbf{LLM-DSE}} \\
\cmidrule(lr){2-2}\cmidrule(lr){3-4}\cmidrule(lr){5-6}\cmidrule(lr){7-8}
& \textbf{\# Cycles} & \textbf{\# Cycles} & \textbf{vs.\,DSE} & \textbf{\# Cycles} & \textbf{vs.\,DSE} & \textbf{\# Cycles} & \textbf{vs.\,DSE} \\
\midrule
3mm            & 189\,570 & 400\,985 & 0.47 & 499\,035 & 0.38 & 26\,539 & \textcolor{darkgreen}{7.14} \\
atax-medium    & 232\,075 & 282\,783 & 0.82 & 279\,591 & 0.83 & 131\,119 & \textcolor{darkgreen}{1.77} \\
covariance     &  29\,668 & 915\,917 & 0.03 & 1\,145\,932 & 0.03 & 29\,344 & \textcolor{darkgreen}{1.01} \\
fdtd-2d        &  25\,054 & --       & --   & 1\,426\,342 & 0.02 & 12\,583 & \textcolor{darkgreen}{1.99} \\
gemm-p         &   9\,179 & 38\,380  & 0.24 & 29\,306  & 0.31 & 9\,120 & \textcolor{darkgreen}{1.01} \\
gemver-medium  & 168\,086 & 262\,431 & 0.64 & 274\,282 & 0.61 & 187\,339 & 0.90 \\
jacobi-2d      & 238\,164 & 4\,335\,041 & 0.06 & 4\,333\,761 & 0.05 & 164\,284 & \textcolor{darkgreen}{1.45} \\
symm-opt       &  13\,277 & 602\,857 & 0.02 & 510\,990 & 0.03 & 13\,277 & 1.00 \\
syr2k          &  51\,581 & 62\,005  & 0.83 & 49\,984  & \textcolor{darkgreen}{1.03} & 22\,740 & \textcolor{darkgreen}{2.27} \\
trmm-opt       &  24\,567 & 477\,490 & 0.05 & 62\,032  & 0.40 & 3\,517  & \textcolor{darkgreen}{6.99} \\
\midrule
\textbf{Geomean} & & & 0.16 & & 0.17 & & \textcolor{darkgreen}{1.87} \\
\textbf{Win \%}  & & & 0\% & & 10\% & & 80\% \\
\bottomrule
\end{tabular}
\end{table*}